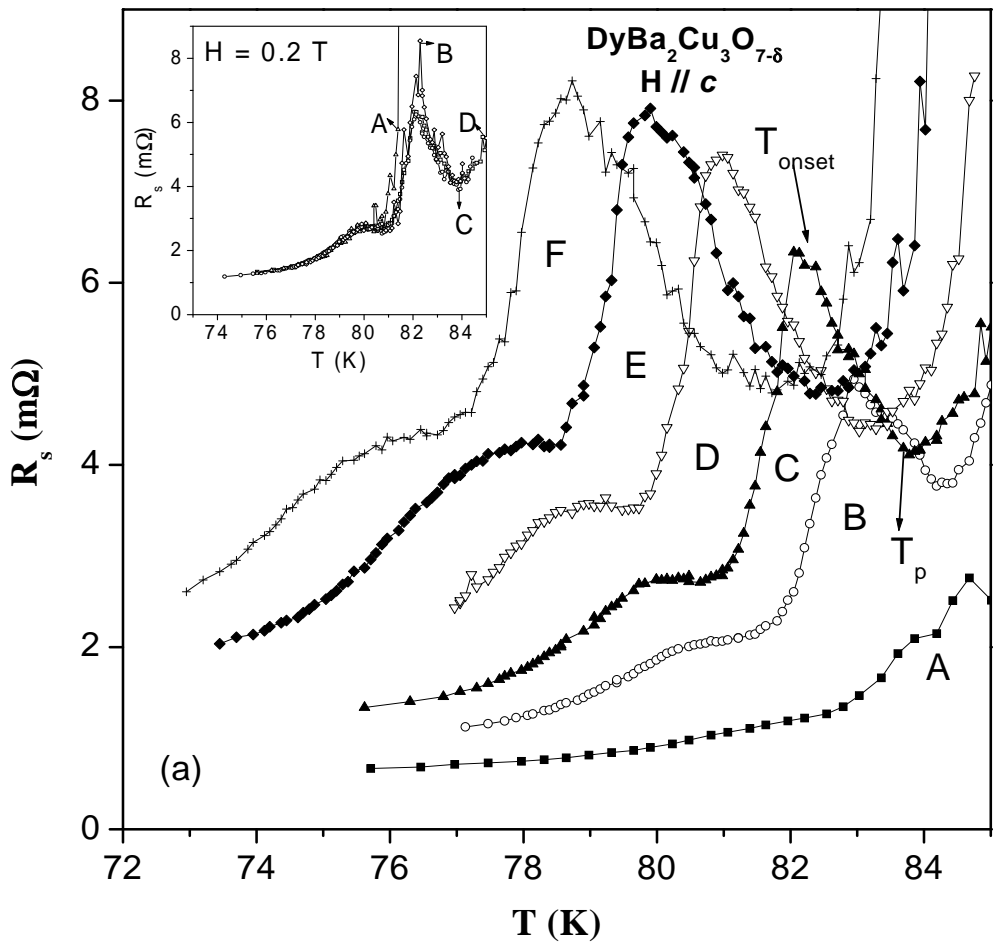

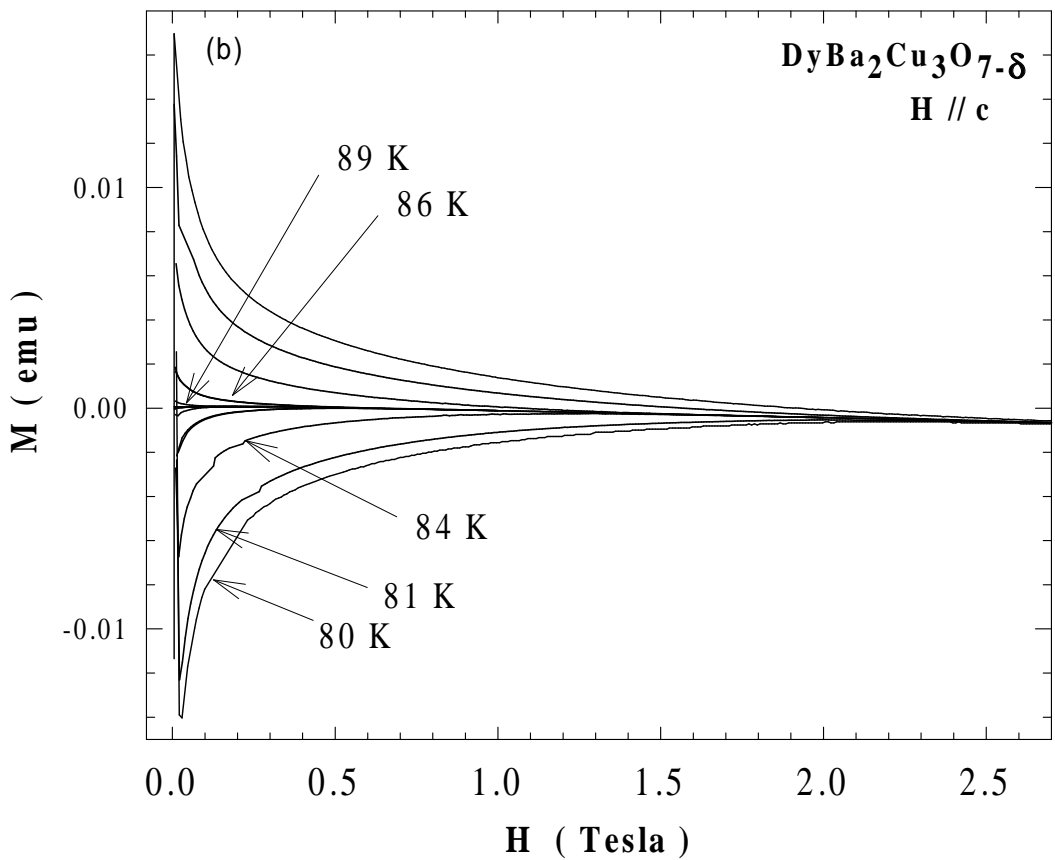

Figure 1 (a) and (b) A.R. Bhangale *et al*

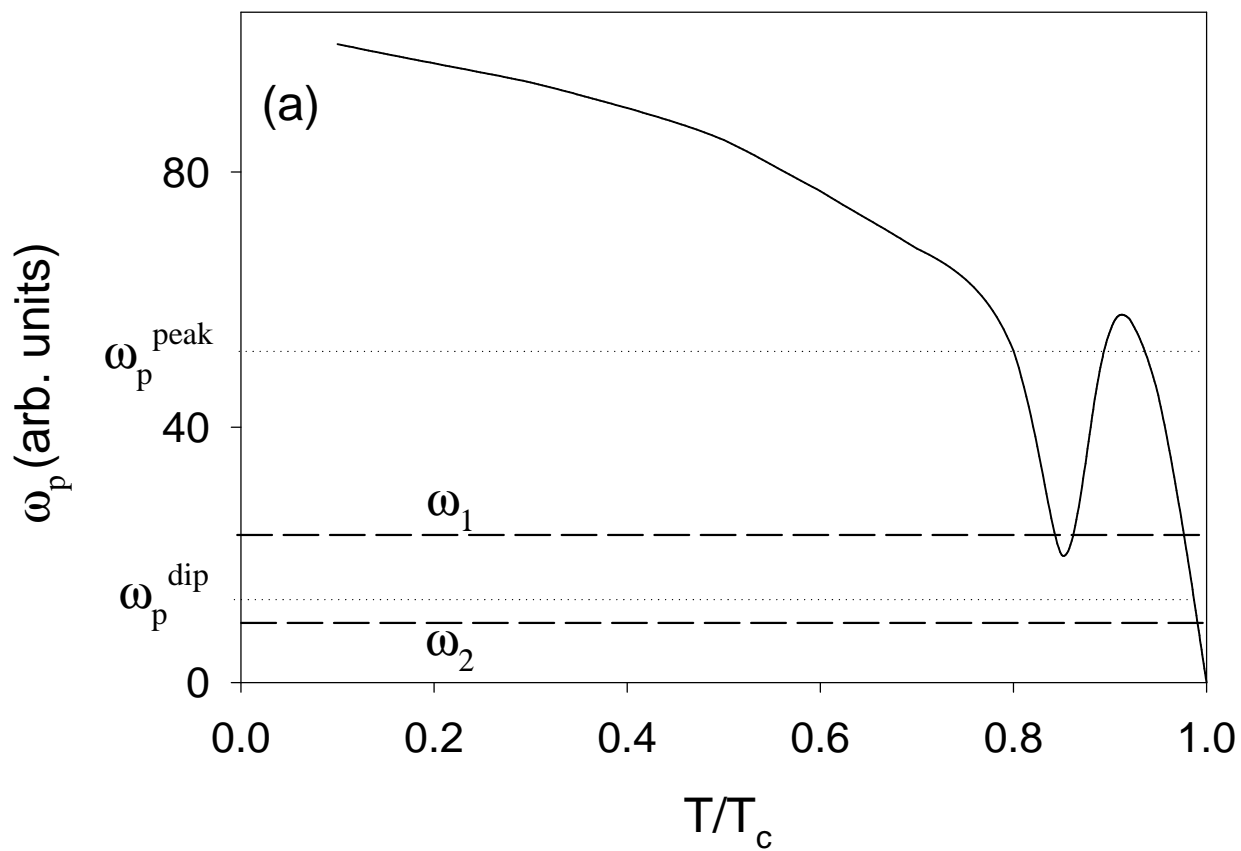

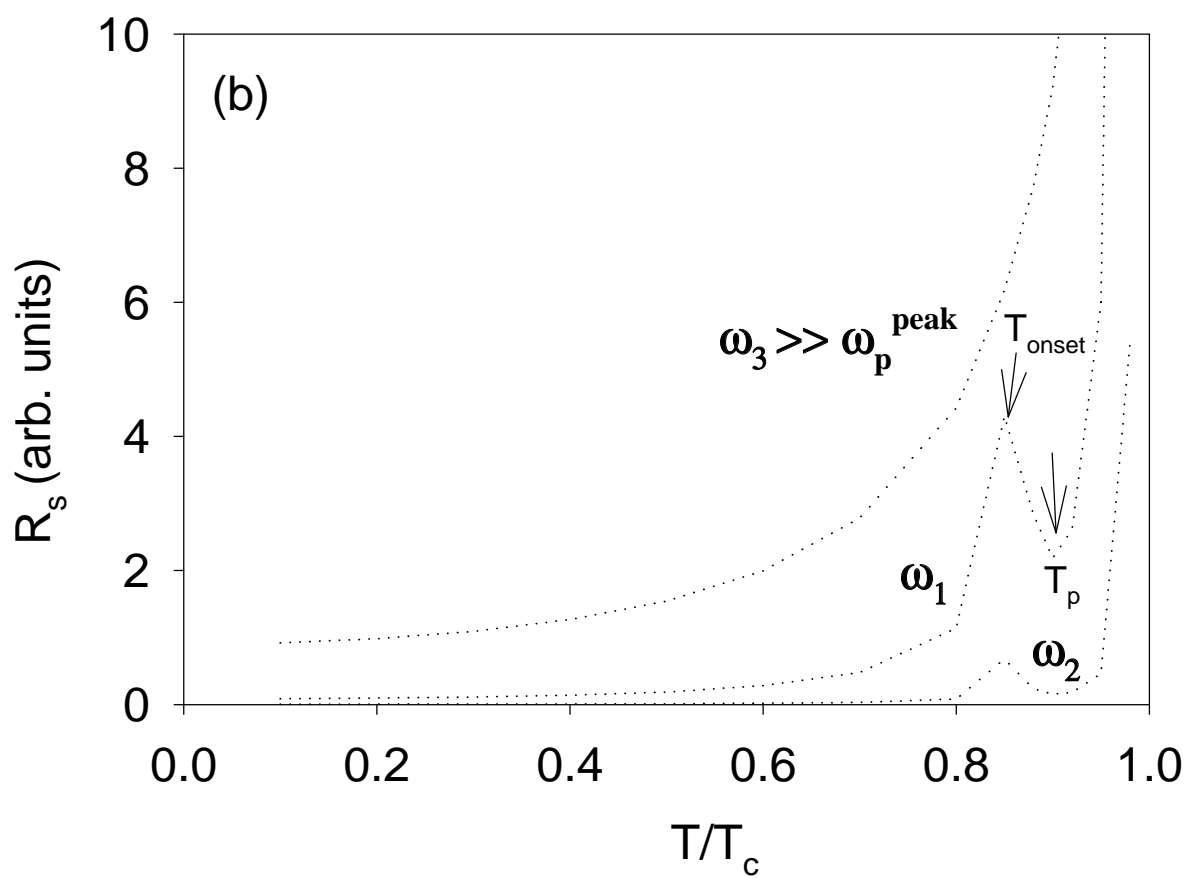

Figure 2 (A. R. Bhangale *et al.*)

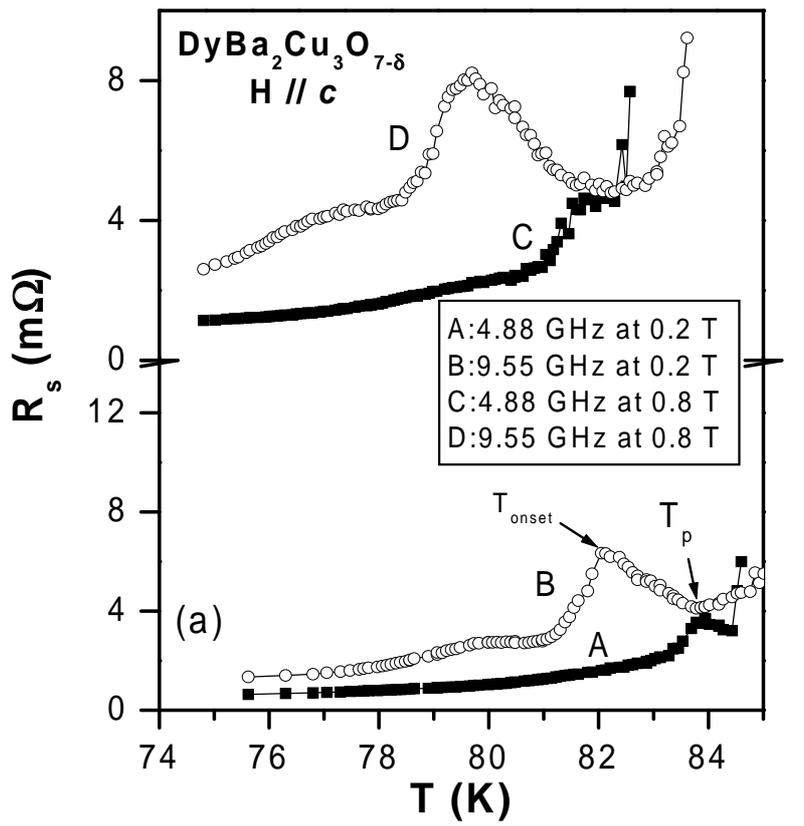

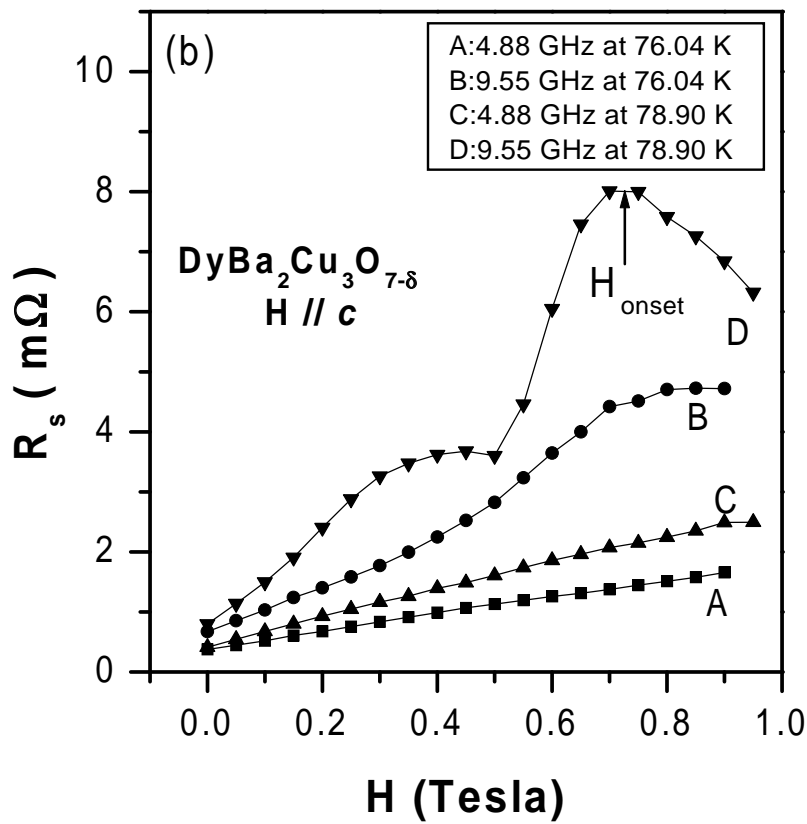

Figure 3 (a) and (b) A.R. Bhangale *et al*

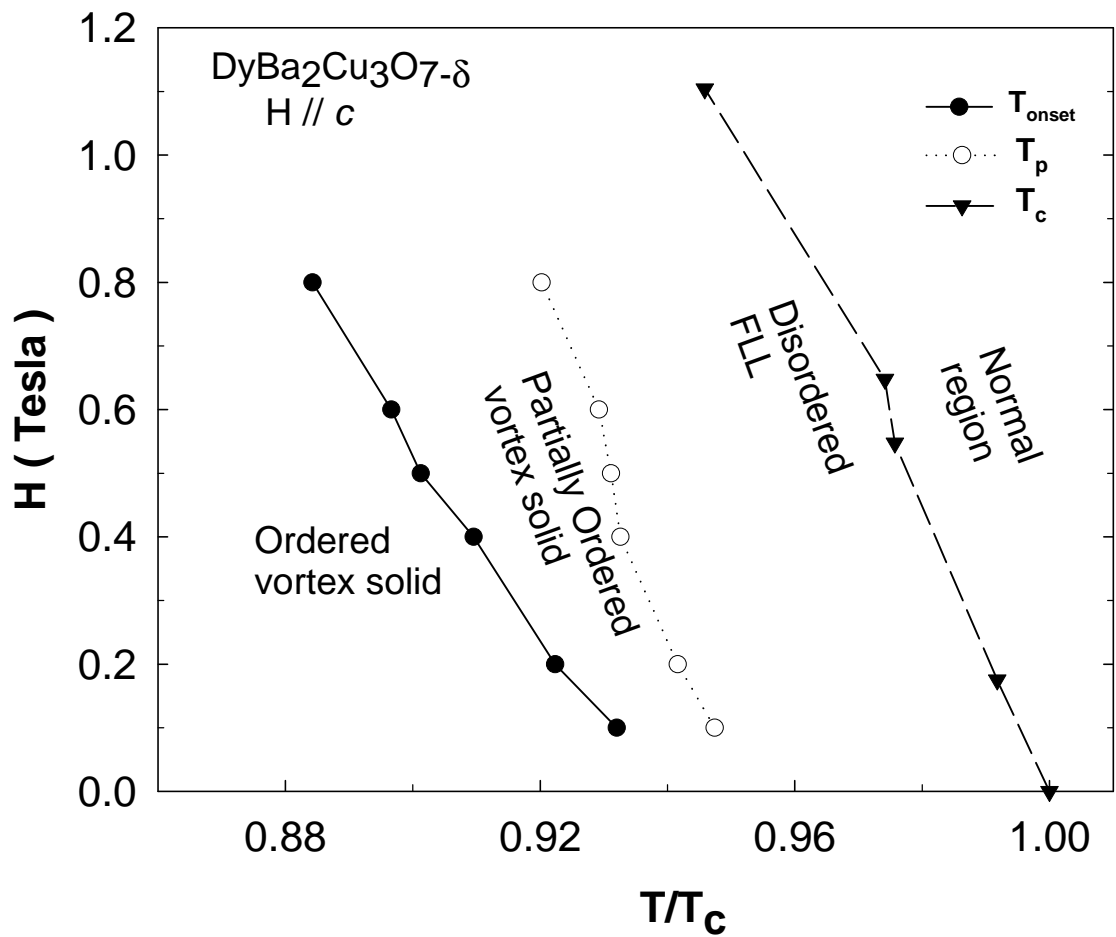

Figure 4 (A.R. Bhangale *et al*)

# Peak effect in a superconducting $DyBa_2Cu_3O_{7-y}$ film at microwave frequencies

**A.R. Bhangale\*, P. Raychaudhuri#[†,1], S. Sarkar[†], T. Banerjee[†],**
**S. S. Bhagwat\*, V. S. Shirodkar\* and R. Pinto[†2]**

[†] *Department of Condensed Matter Physics and Materials Science, Tata Institute of Fundamental Research, Homi Bhabha Rd., Colaba, Mumbai 400005, India.*

[#] *School of Physics and Astronomy, University of Birmingham, Edgbaston, Birmingham, B15 2TT, UK.*

[\*] *Department of Physics, Institute of Science, 15 Madam Cama Rd., Mumbai 400032, India.*

*Abstract:* We report the observation of a peak in the microwave ( 9.55 GHz ) surface resistance in an epitaxial $DyBa_2Cu_3O_{7-y}$ superconducting film in magnetic fields (parallel to the *c* axis) ranging between 0.2 to 0.9 Tesla. Such a peak is absent in the measurements done in zero-field. The temperature and field dependence of the peak suggests that this peak could be associated with the peak effect phenomenon reflecting the order-disorder transformation in the flux-line lattice. A strong dependence of this peak effect at frequencies close to the depinning frequency of the flux line lattice is observed.

---

[1] e-mail: P.Raychaudhri@bham.ac.uk
[2] e-mail: rpinto@tifr.res.in

The phenomenon of peak-effect (PE) in type II superconductors has attracted widespread attention in order to understand the order-disorder transition[1] in the flux line lattice (FLL) in the mixed state of type II superconductors. The physical phenomenon associated with the peak effect is the occurrence of a peak in the critical current density $J_c$ below its superconducting-normal phase boundary[2]. In a varying temperature measurement, the vortex state in a type II superconductor undergoes a transition from an ordered state below the peak temperature ($T_p$) to a highly disordered state above $T_p$[3-7]. This phenomenon is rationalized within the Larkin-Ovchinnikov[8] scenario, where the effective pinning force on the FLL is given by the expression,

$$BJ_c(H) = (n_p<f^2>/V_c)^{1/2}, \qquad (1)$$

where $n_p$ is the density of pinning sites, $f$ is the elementary pinning force parameter, B is the magnetic induction and $V_c$ is the volume of Larkin domain within which the vortex lattice retains its spatial order. At $T_p$, the $V_c$ reaches a minimum value due to the disordering of the FLL, thereby enhancing the critical current. Though this general description is widely accepted, the exact nature of the order-disorder transformation is still an issue of intense research. Several scenarios, such as the melting of the Larkin domains to a vortex liquid state[1,4] or fracturing of an ordered vortex solid to an amorphous state in the presence of random quenched disorder[7,9], have been proposed as the physical picture governing the peak effect phenomenon in both low $T_c$ and high $T_c$ superconductors.

In recent years, the statics and dynamics of the FLL has been probed through various transport, magnetic, and structural measurement techniques. Whereas the static properties of the FLL are measured through steady state measurements, such as, dc magnetization and Bitter decoration, the dynamics of the vortex state is probed either by applying a small perturbation and measuring the response of the vortex to the perturbation or via transport measurements above the critical current ($J_c$). Towards the latter end, the ac susceptibility measurements[3,9,10] have been extensively carried out to understand the dynamics of FLL. These measurements are typically carried out with an ac excitation field in the range of few tens of Hz to few MHz. These measurements reveal no frequency dependence of the peak position of the PE, which suggest the association of a true thermodynamic phase transition with this effect[10]. All these measurements however, probe the variation of the critical current in the superconductor, where the force on the vortices becomes equal to the maximum pinning force. On the other hand, a small microwave excitation induces a current, which is smaller than the critical current. Therefore, the vortices move back and forth close to the minimum of the pinning potential and will experience the restoring force close to the potential minimum. There have been few studies[11–15] on the dynamics of the FLL in the microwave and radio frequency regime. The dynamics of the vortices at these frequencies can be described by an equation of motion suggested by Gittleman and Rosenblum[16] and given as,

$$\eta \dot{x} + kx = F, \qquad (2)$$

where $\eta$ is the Bardeen-Stephen viscous drag coefficient, $k$ is the pinning constant and F is the external force on the vortex given by, $F = J\Phi_0$, where $\Phi_0$ is flux quantum $hc/2e$. It can be easily shown that the vortex impedance is given by,

$$\rho_v = \frac{\Phi_0 H}{\eta \left(1 + i \frac{\omega_p}{\omega}\right)}, \tag{3}$$

where $\omega_p = (k/\eta)$ is the depinning frequency. At low frequencies ($\omega \ll \omega_p$), the vortex impedance is mostly inductive and dominated by pinning; at high frequencies ($\omega \gg \omega_p$), the motion is mostly resistive with pinning playing a very minor role. From the theoretical point of view, the dynamics of the FLL at high frequencies is interesting since at such frequencies ($\omega \gg \omega_p$), the FLL is expected to behave as if it is unpinned[16]. Flux flow behavior of the FLL at microwave frequencies has indeed been observed in both low $T_c$ and high $T_c$ superconductors[11,17]. Till date, there is no report pertaining to the observation of the peak effect at microwave frequencies in either a low $T_c$ or a high $T_c$ superconductor. It is however instructive to explore the applicability of eqn. 2, within the collective pinning scenario[8,16], where the vortices within a Larkin volume elastically respond like a semi-rigid body. In such a case, the total external force per unit volume on the FLL within $V_c$ is given as, $F = n\Phi_0 J = BJ$ (where n = vortex density). On the other hand, the total restoring force per unit volume will be same as in eqn. 1 and, therefore, $k \propto (n_p \langle f^2 \rangle / V_c)^{1/2}$. This will have the same temperature and field variation as $J_c$, and will show a peak like feature close to $T_c$ (or $H_{c2}$). This should therefore give a minimum in the surface resistance $R_s$ at the order-disorder transition of the vortex lattice, where $V_c$ reaches the minimum limit.

In this paper we report the observation of a pronounced peak effect at microwave frequencies in a DyBa$_2$Cu$_3$O$_{7-\delta}$ (DBCO) superconducting film (2500Å) grown by pulsed laser deposition on a single crystalline LaAlO$_3$ substrate. The film was initially characterized by ac susceptibility measurements using the flat coil geometry and its $T_c$ was found to be 90±0.2 K . The film was subsequently patterned into a linear stripline with a width of 175 µm and length 10 mm, and the measurements of microwave absorption were performed by stripline resonator technique using a scalar network analyzer Model HP8757, synthesized sweeper Model HP83620A and a closed cycle helium cryocooler. Details of this measurement have been described elsewhere[18]. DC magnetic field up to 0.9 Tesla was applied perpendicular to the film plane (parallel to the c-axis of DBCO film) using an electromagnet.  The temperature fluctuation during an isothermal measurement remained within 30 mK.

Figure 1(a) shows the temperature variation of the surface resistance ($R_s$) at 9.55 GHz (corresponding to the first harmonic excitation of the stripline) measured in various magnetic fields. In all these measurements, the current induced by the microwaves was much smaller than the critical current of the material. Note that $R_s$ displays a pronounced maximum followed by a dip feature before the superconducting transition temperature ($T_c$). The temperatures corresponding to extreme positions shift to lower values as the magnetic field is increased. The inset shows the plots of $R_s$ measured at different microwave power levels. It is apparent that the temperatures corresponding to the characteristic features of the anomalous variation in $R_s$ do not change with the variation in the microwave power level, indicating that the currents induced by the microwave field are lower than the critical current of the superconductor. One disadvantage of the stripline resonator technique is that one cannot access the normal state $R_s$ of the superconducting material, and, hence, cannot determine the $T_c(H)$ very accurately. The $T_c(H)$ was therefore estimated from the upper critical field ($H_{c2}$) determined from isothermal magnetization versus field (M-H) measurements [see Fig. 1(b)] on another film grown under identical conditions. We could not observe any signature of the peak effect phenomenon in the M-H loops in our thin film sample. A possible reason for this behavior will be elaborated later on. In Fig. 3(b) we have shown the variation of $R_s$ with applied magnetic field at various temperatures. Here, also a maximum is observed in the $R_s(H)$ at 9.55 GHz, which coincides with parametric values in the H-T plane, with the maximum observed in $R_s(T)$ [at $T_{onset}$] during temperature sweep measurements. It is however interesting to note that the peak in the surface resistance is not observed in the $R_s$ measured from the fundamental resonance (4.88 GHz). Unfortunately, the field range available in our experiment does not allow us to probe the peak effect at temperatures lower than 76 K.

To understand the origin of the peak in $R_s$, we have to consider the evolution of pinning constant, k, [cf. eqn. (2) ] within the collective pinning scenario. The evolution of k is similar to that of $J_c$ and will show a peak at the order-disorder transition (where $V_c$ goes to a minimum value) as the field or temperature is increased. Since within the Bardeen-Stephen model the viscosity $\eta$ varies smoothly with temperature, the depinning frequency $\omega_p$ will also show a minimum followed by a peak at the order-disorder transition [see the schematic drawn in Fig. 2(a)]. We identify the frequency at the dip as $\omega_p^{dip}$ and at the peak as $\omega_p^{peak}$, respectively. The observation of a peak in the $R_s$ will critically depend on the measurement frequency. When $\omega \gg \omega_p^{peak}$, the $R_s$ will increase monotonically with field or temperature [see the schematics in Fig. 2(b)] without showing any peak. When $\omega_p^{peak} > \omega > \omega_p^{dip}$ [i.e., $\omega_1$ in Fig. 2(b)], the measurement frequency will become larger than the depinning frequency at some temperature and the $R_s$ will increase.

*Bhangale et al.*

However, since the depinning frequency passes through a peak, at a higher temperature the measurement frequency will again become lower than the depinning frequency causing the $R_s$ to decrease. Therefore, in this frequency range, the $R_s$ will show a pronounced peak [cf. Fig. 2(b)]. The position of the peak and the subsequent dip in $R_s$ will coincide with dip and the peak in $\omega_p$, respectively. This also follows from the eqn. 3. On the other hand, when $\omega < \omega_p^{dip}$ [i.e., $\omega_2$ in Fig. 2(b)], the peak will be less pronounced since the measurement frequency will cross the depinning frequency only once, where the surface impedance will undergo a cross-over from predominantly inductive to predominantly resistive behaviour and $R_s$ will increase. Usual estimates of the depinning frequency in cuprate superconductor, like, $YBa_2Cu_3O_{7-\delta}$ vary between 5 to 40 GHz[11]. This is therefore consistent with the fact that we observe a pronounced peak when the measurements are done at 9.55 GHz. The main difference between this microwave peak effect and the peak effect observed in conventional low frequency measurements, such as, magnetization or transport critical current experiments is that in low frequency measurements one actually measures the peak in the critical current density. This corresponds to the point where the force on the vortex lattice is equal to the maximum pinning force $(dV(x)/dx)_{max}$ and the vortices start moving out of their pinning potential. Therefore, most conventional measurements probe the temperature or field dependence of $(dV(x)/dx)_{max}$. In microwave experiments, however, the current is lower than the critical current and therefore a vortex moves close to its potential minimum. The microwave measurements hence probe the variation in $(dV(x)/dx)_{x\approx 0}$. It is however to be noted that equation (2) is strictly not applicable at temperatures close to $T_c$ due to large-scale motion of the vortices. This point will be elaborated further.

To explore the frequency dependence of $R_s$, the surface resistance was also measured using the fundamental resonance (4.88 GHz) of the stripline. The plots of $R_s$ as a function of temperature at 4.88 GHz and 9.55 GHz are shown in Fig. 3(a) at magnetic fields of 0.2 Tesla and 0.8 Tesla, respectively. $R_s$ as a function of magnetic field for the same frequencies are shown in Fig. 3(b). Both for the isothermal and the isofield runs, we observe that the peak effect is much less pronounced and is absent for temperatures below 80 K [Fig. 3(b)], when the measurements are carried out at a frequency of 4.88 GHz. This is consistent with the scenario proposed earlier. However, one major discrepancy with the earlier scenario is the shift in the peak position to higher temperatures when the measurements are done at 4.88 GHz. According to eqn. (3) the parametric values of the peak and the dip positions in the $R_s$ should be independent of the frequency. The usual treatment of the frequency dependence of the Labusch parameter due to thermal creep using the independent vortex scenario is also unable to account for the frequency dependence of the dip in the $R_s$ values. This discrepancy possibly arises due to the limitations in our assumption that the collective pinning description strictly holds all across the PE region.

The independent vortex picture of the Gittleman and Rosenblum[16] described by eqn. 3 assumes that each vortex remains within the pinning potential minimum and therefore the process of flux creep does not get taken into account. In the collective pinning scenario[8], this picture remains valid provided the motion of the vortices inside $V_c$ is small compared to the overall pinning potential arising collectively from the pinning centers inside the Larkin domain. Thus, within $V_c$, the vortices do not experience the distribution in the restoring force arising from the distribution in the pinning potential in the system. However, close to the order-disorder transition, the usual collective pinning scenario[8] may require some modification as the vortex state transform to an amorphous phase[7, 9] where individual vortices (or bunch of vortices) are pinned in potentials of varying strength *k*. The evidence for this kind of glassy state close to the PE in a twinned $YBa_2Cu_3O_7$ crystal (with H // c) has recently been observed by Pal *et al*[19]. This could cause the system to have a distribution in time scales and become the plausible cause for the shift in position of the peak in $R_s$, when the measurements are carried out at different frequencies. At present we do not, however, have a more precise model to reproduce the observations.

A point worth considering while analyzing the response of vortices at high frequencies is the effect of surface pinning. Placais and co-workers[14,15] have analyzed the effect of surface pinning on the vortex response in a mixed state of type-II superconductors by introducing the line tension of the vortex held by surface pinning. By studying the frequency response of the superconductor in the mixed state, they[15] argue that the vortices in a single crystal of YBCO are held by surface pinning, while they are free to move inside the bulk. These experiments arguably question the collective pinning scenario, where a uniform density of pins is assumed in the bulk. Their experiments were however restricted to a narrow temperature range close to $T_c$, which is higher than the range of temperature where we have observed a non-monotonic behaviour in $R_s$. It is not known, at this stage, how the collective pinning scenario can be modified in the presence of surface pinning alone. However, there are two factors due to which bulk pinning might be responsible for the observed behaviour in our sample. Firstly, the epitaxial film grown by laser ablation has a larger density of defects, including extended defects such as twin boundaries as compared to high quality single crystals which are very weakly pinned. Such defects have been observed in these thin films with atomic force microscopy. These extended defects act like surfaces inside the bulk of the crystal. Secondly, the thickness of these films are of the order of the penetration depth in these superconductors near the measurement temperature, which makes the distinction between surface and bulk pinning less significant as well as difficult to detect. However, this issue can be resolved by a complete spectral analysis of the complex penetration depth over a wide frequency range, which is beyond the scope of the present study.

Based on the temperature variation of $R_s$ at 9.55 GHz measured at various fields, we have constructed in Fig. 4 a tentative vortex phase diagram for $DyBa_2Cu_3O_{7-\delta}$ (H // c). The peak in $R_s(T)$, which corresponds to a minimum in the depinning frequency marks the onset temperature ($T_{onset}$) of the order-disorder transformation in the vortex state. The process of disordering gets completed at the peak of $\omega_p$ which corresponds to the minimum of $R_s$. This temperature is denoted as $T_p$ (cf. Fig. 1). The phase diagram comprises an ordered vortex state, which crosses over to a fully disordered state via a partially ordered phase, as the temperature or field is increased. This is in agreement with the vortex phase diagrams proposed at lower frequencies[9]. However, an additional frequency dependence of the microwave peak effect makes it interesting to study and establish the dynamics of the system over a larger frequency range.

One last thing to ponder about is the absence of any fingerprint of the peak effect in our dc magnetization measurements. One of the differences between magnetization and microwave measurements on epitaxial thin films is that in magnetization measurements one actually probes the macroscopic shielding currents (intergranular as well as intragrain) induced in the superconductor. The microwave radiation on the other hand will induce small current loops, which see a much smaller cross-section of intergranular weak links as compared to the currents induced in magnetization measurements. This could be the possible reason for the absence of peak effect in magnetization data on any of the laser ablated $YBa_2Cu_3O_{7-\delta}$ and $DyBa_2Cu_3O_{7-\delta}$ thin films reported so far.

In conclusion, we have observed a pronounced peak effect in a thin film of $DyBa_2Cu_3O_{7-\delta}$ at subcritical currents at microwave frequencies close to the depinning limit (9.55 GHz) of the superconductor. This peak effect has been attributed to the order-disorder transformations of the FLL as the temperature or field is increased. In contrast to the low frequency measurements on the PE, this phenomenon at microwave region has pronounced frequency dependence both in terms of the magnitude of the effect as well as position of the peak temperature. It would be interesting to study PE in high temperature superconductors over a wider frequency range and at higher fields to understand the interrelation between the Labusch parameter, Larkin volume, and the vortex viscosity and their effect on the order-disorder transition of the FLL.

*Acknowledgements:* The authors would like to thank Professor A.K. Grover, Professor Shobo Bhattacharya and Professor S. Ramkrishnan for useful discussions.

**FIGURE CAPTIONS**

**Fig. 1:** (a) Surface resistance $R_s$ at a frequency of 9.55 GHz with 10 dB power for various applied fields (parallel to *c*) as a function of temperature. Here A = 0, B = 0.1, C = 0.2, D = 0.4, E = 0.6 and F = 0.8 Tesla. Inset: $R_s$ vs T plot in H = 0.2 Tesla at various power levels, A = 0, B = 2, C = 5 and D = 10 dB. (b) Isothermal M-H loops between 80 and 89 K as indicated.

**Fig. 2 :** Schematic diagrams of (a) variation of depinning frequency $\omega_p$ with reduced temperature, which shows a dip followed by a peak in the order-disorder transition region, and (b) surface resistance $R_s$ vs temperature. Note that the comparison of plots in (a) and (b) shows that since $\omega_p^{peak} > \omega_1 > \omega_p^{dip}$, $R_s$ shows a pronounced peak and a dip, whereas for $\omega_2 < \omega_p^{dip}$, $R_s$ shows a less pronounced peak. For $\omega_3 >> \omega_p^{peak}$, the peak in $R_s$ is not visible.

**Fig. 3:** (a) Plot of $R_s$ vs temperature measured at fields (// to *c*) at 0.2 and 0.8 Tesla and with frequencies of 4.88 GHz and 9.55 GHz. (b) $R_s$ as a function of magnetic field at different temperatures and frequencies.

**Fig. 4:** Vortex phase diagram in a $DyBa_2Cu_3O_7$ sample for H // *c*.